\documentclass[12pt]{article}

\usepackage{graphicx}
\usepackage{color}

\begin{document}
 \title{Numerical Study of Pseudogap Anisotropy
 for a Tight Binding Band}
 \author{L. Coffey,\\
 Physics Department,\\ 
 Illinois Institute of Technology,\\
 Chicago, Illinois 60616}
\maketitle
\begin{center}
The Fermi surface anisotropy of a pseudogap in the spectral weight, caused
by superconducting or antiferromagnetic fluctuations, is calculated for the
case of a tight binding band. The importance of the saddle point, or {\em hot
spot}, on the Fermi surface is illustrated. A peak-dip-hump structure in
the spectral weight, due to a finite frequency mode, is also briefly explored.
The results are applied to high temperature cuprate superconductors.

\end{center}

\newpage

{\bf Introduction}\\
\\
Among the properties of the high temperature superconducting cuprates, that have yet to be fully explained, are  the Fermi surface anisotropy of the pseudogap \cite{1} in the normal state, and that of the peak-dip-hump feature \cite{1a} in the superconducting state, both of which are seen in ARPES measurements of the electronic spectral weight.
These features are maximal near the saddle point of the tight binding band that describes these materials, and
disappear gradually for Fermi surface momenta away from that point.

The purpose of this paper is to focus on this one aspect of the complicated behavior and origins of the pseudogap, and to
show how the anisotropy can arise solely from the properties of the tight binding band.

The saddle point is also where the cuprate d-wave superconducting gap is at its maximum, or antinode, 
and this has led to much speculation concerning the origin
of the pseudogap, including the possibility of preformed Cooper pairs in the normal state. 

Other possible causes of the pseudogap are superconducting fluctuations or
antiferromagnetic spin fluctuations. There is also evidence of a pseudogap arising 
from precursor superconducting pairing,
and a competing mechanism, operating simultaneously in the normal state above T$_{C}$
\cite{2}.
A numerical study is presented here of
pseudogaps resulting from these types of mechanisms in the
spectral weight, and the accompanying density of states, using a tight binding band that is relevant for the cuprates. The results presented here are a follow-up on previous work on superconducting fluctuations in a tight binding band \cite{3}.

One feature of the tight binding band,
the possibility of the Fermi surface being close to the saddle point region in the Brillouin
zone, leads on its own to an anisotropic pseudogap on the Fermi surface
The importance of the saddle points in modelling the electronic properties
of the cuprates has led to the term {\em hot spots} being coined to denote them \cite{4}.   
Experimental evidence for an {\em extended saddle point singularity} has been
observed in ARPES measurements on B-2212 \cite{4a}. The absence of a corresponding Van Hove type peak in the
experimenal tunneling density of states can be due to the role of tunneling directionality \cite{4b}.

Furthermore, the degree of nesting in the Fermi surface plays a significant role in determining the pseudogap behavior due to antiferromagnetic spin fluctuations. For a Fermi surface without nesting, such as is the case in optimally doped B-2212, a pseudogap in the 
electronic spectral weight may exist over a region of the Fermi
surface without resulting in a pseudogap in the accompanying density of
states.

In the previous work \cite{3}, the superconducting fluctuation propagator
was calculated in order to connect quantitatively the magnitude of the
fluctuation self energy to values for T$_{C}/t$ which are typical for a cuprate superconductor. In the results presented here, standard phenomenological models are used 
for the fluctuation propagators for antiferromagnetic spin fluctuations and
superconducting fluctuations. In the former case, an antiferromagnetic fluctuation propagator with a peak wavevector 
$\vec{q }=(\pi,\pi)$ \cite{4} is used, and in the latter case, a propagator peaked about $q =0$ and $\omega=0$ \cite{5}.

A peak-dip-hump structure appears in some cuprates at the {\em hot spot} in
the Fermi surface spectral weight, measured by ARPES, below the superconducting transition temperature T$_{C}$
Its origin is attributed to a $\vec{q} =(\pi,\pi)$ mode,
which emerges below T$_{C}$,
with energy in the same range as the value of the
superconducting gap \cite{6}. The effect of a finite frequency ($\pi$,$\pi$) mode on the normal state electronic spectral weight is investigated in this study. A peak-dip-hump
structure results in the spectral weight at the {\em hot spot}, and is absent for the Fermi surface momentum where the nodal point of the d-wave superconducting gap is located
While the present study deals with the
normal state spectral weight, the results of this work suggest that  proximity of the cuprate Fermi surface to the {\em hot spot} 
is a likely explanation for the anisotropy of both the peak-dip-hump feature and the pseudogap. \\
\\
\\

{\bf Theory}\\
\\
\\
The effects of fluctuations, or of a finite frequency mode, are incorporated into the spectral weight 
$A( \vec{p}, \omega )$, and density of states $N(\omega)$, through the self-energy $ \Sigma(\vec{p}, \omega )$ 
with
\begin{equation}
A( \vec{p}, \omega )\; = \; - \frac{1}{\pi} \frac{ {\rm Im} \Sigma(\vec{p}, \omega) }{
(\omega - \varepsilon_{p} - {\rm Re} \Sigma(\vec{p}, \omega ))^{2} + ({\rm Im} \Sigma(\vec{p}, \omega))^{2} }
\end{equation}
and
\begin{equation}
N(\omega)\; = \; \frac{1}{(2 \pi)^2} \int_{- \pi} ^{\pi} dp_{x}\int_{- \pi} ^{\pi}{\rm d}p_{y} \;
A( \vec{p}, \omega )
\end{equation}
The tight binding band is given by
$\varepsilon_{p}= -2t( {\rm cos}(p_{x})+{\rm cos}(p_{y}))-4t^{'} {\rm cos}(p_{x}) {\rm cos}(p_{y})-\mu$.\\
\\
$\Sigma(\vec{p}, \omega)$ represents the self-energy due to one of the three processes investigated 
in the present work: normal state superconducting fluctuations, antiferromagnetic spin fluctuations with dominant wavevector
$\vec{q}=(\pi,\pi)$, and a mode with finite frequency $\Omega_{0}$ and dominant wavevector $\vec{q}=(\pi,\pi)$.\\
\\
The superconducting fluctuation self energy is defined \cite{7} as
\begin{equation}
\Sigma_{SC}(\vec{p}, i \omega_{n} ) \; = \; - T\Sigma_{q,i \omega_{m}} G(\vec{q}- \vec{p}, i \omega_{m}-i \omega_{n}) L_{SC}(\vec{q},i\omega_{m})
\end{equation}
where $L_{SC}(\vec{q},\omega^{'})$ is the superconducting fluctuation propagator.
This yields 
\begin{equation}
\Sigma_{SC}(\vec{p}, \omega) \; = \; \int \frac{d^{2} q}{(2 \pi)^{2}}  \int 
\frac{d \omega^{'}}{\pi} 
{\rm Im}L_{SC}(q, \omega^{'})
\frac{[ n_{B}( \omega^{'}) + n_{F}( \varepsilon_{q-p})] }{ 
\omega^{'} - \omega -\varepsilon_{q-p} - i \delta }
\end{equation}
where $n_{B}( \omega^{'})$ and $n_{F}( \varepsilon_{q-p})$ are the Bose-Einstein
and Fermi-Dirac distributions.\\
\\
The superconducting fluctuation propagator $L_{SC}(q,\omega)$ is given by \cite{5}
\begin{equation}
L_{SC}(q,\omega) \; = \; - \frac{\alpha_{SC}}{1+(\zeta_{SC} q)^{2} - i \omega/ \omega_{SC}}
\end{equation}
where $\alpha_{SC}$, 
$\zeta_{SC}$, and $\omega_{SC}$ are a superconducting coupling constant, the correlation length for superconducting fluctuations, and a characteristic superconducting fluctuation
frequency respectively.\\
\\
The self energy for antiferromagnetic spin fluctuations, or due to a finite frequency mode, is calculated with \cite{8}
\begin{equation}
\Sigma_{AF/Mode}(\vec{p}, i \omega_{n}) \; = \; - T \Sigma_{q,i \omega_{m}} G(\vec{p}- \vec{q}, i \omega_{n}-i \omega_{m})
L_{AF/Mode}(\vec{q},\omega_{m})
\end{equation}
which results in
\begin{equation}
\Sigma_{AF/Mode}(\vec{p}, \omega) \; = \; \int \frac{d^{2} q}{(2 \pi)^{2}}  \int 
\frac{d \omega^{'}}{\pi} 
{\rm Im}L_{AF/Mode}(q, \omega^{'})
\frac{[{\rm coth}( \omega^{'}/2T) + {\rm tanh}( \varepsilon_{p-q} /2T) ]}{ 
\omega^{'} - \omega +\varepsilon_{p-q} - i \delta }
\end{equation}
\\
The antiferromagnetic spin fluctuation propagator $L_{AF}(q, \omega)$ is modelled using
\cite{4}
\begin{equation}
L_{AF}(q,\omega) \; =  \; - \Sigma \Sigma_{s,s^{'}= \pm 1}\frac{\alpha_{AF}}{1+(\zeta_{AF} Q)^{2} - i \omega/ \omega_{AF}}
\end{equation}
which is the same functional form as equation (5), with 
$\alpha_{AF}$, $\zeta_{AF}$, and $\omega_{AF}$, a magnetic coupling constant, the correlation length for spin fluctuations, and a spin fluctuation
frequency respectively. The spin fluctuation propagator $L_{AF}(q,\omega)$ is peaked about $\vec{q} =  (\pi,\pi)$ with
 $Q^{2} \; = \; (q_{x} + s \pi)^{2}+(q_{y}+ s^{'} \pi)^{2}$ in equation (8).\\

The effect on the spectral weight of a $\vec{q}=(\pi,\pi)$ mode, with frequency
$\Omega_{0}$, is calculated using equation (7) with Im$L_{Mode}(q,\omega)$ defined by
\begin{equation}
{\rm Im} L_{Mode}(q,\omega) \; = \; - 
c_{Mode}^{2} \; A \; \omega \;
 {\rm exp} [-  \frac{(|\omega| - \Omega_{0})^{2})}{\Delta^{2}}] \; \; \Sigma \Sigma_{s,s^{'}= \pm 1} \frac{1}{[(1.0+Q^{2})^{2}]}
\end{equation}
where $c_{Mode}$ is the mode coupling constant, $A$ the peak magnitude of the mode spectral weight, $\Omega_{0}$ the mode
frequency, $\Delta$ the peak width, and $Q^{2} \; = \; (q_{x} + s \pi)^{2}+(q_{y}+ s^{'} \pi)^{2}$.\\
\\
{\bf Results}\\
\\
Typical results for the real and imaginary parts of the different self energies $\Sigma(\vec{p}, \omega )$ (in units of the hopping parameter $t$) are shown in Figure 1
for superconducting fluctuations, antiferromagnetic fluctuations, and the finite frequency mode at the Fermi surface momentum closest to the {\em hot spot} or $\vec{p}=(\pi,0)$ with $t^{'}=-0.35t$ and $\mu=-1.4t$ in $\varepsilon_{p}$.
For Fermi surface momenta away from the {\em hot spot}, the magnitude of these self energy
components decrease monotonically for all three cases. \\
\\
The parameters values used 
are $\alpha_{SC}=46t, \; \zeta_{SC} = \sqrt(10) a \; $ ($a$  is the lattice constant),$ \; \omega_{SC}=t/25$ in eqn.(5),
$\alpha_{AF}=25t, \; \zeta_{AF} = \sqrt(10) a, \; \omega_{AF}=t/20$ in eqn.(8), and $c_{Mode} = 1 {\rm eV}, \; t=0.15 {\rm eV}, \;
A= 66 {\rm eV}^{-2}, \; \Omega =0.35t, \; \Delta = 0.1t$ in eqn. (9). The temperature
is fixed at $T=0.17t$ for all calculations.\\
\\
Figure 2 shows a set of spectral weight $A( \vec{p}, \omega )$ curves incorporating
normal state superconducting fluctuations (left side column in Figure 2, eqns.(4) and (5)), and
antiferromagnetic spin fluctuations (right side column in Figure 2, eqns. (7) and (8)) for a set of Fermi surface momenta $\vec{p}$ starting at the
{\em hot spot} in the top panel and moving away along the Fermi surface for 
$t^{'}=-0.35t$ and $\mu=-1.4t$ in $\varepsilon_{p}$
The $\vec{p}$ values are marked on the
left side column of the panels in Figure 2. The right side column uses the same $\vec{p}$ values.

The bare chemical potential $\mu$ is
replaced with $\mu+ {\rm Re} \Sigma( \vec{p}, \omega = 0)$ when plotting the spectral weight
$A( \vec{p}, \omega )$.
The resulting spectral weights satisfy the sum rule 
$\int {\rm d} \omega A( \vec{p}, \omega ) \; = \; 1$ to within a few percent accuracy, and are
almost symmetric about $\omega=0$.\\
\\
Figure 2 illustrates how proximity to the {\em hot spot}
 results in a pseudogap in $A( \vec{p}, \omega )$, which disappears
at a Fermi surface momentum $\vec{p}$  before the nodal point of the d-wave superconducting state gap is reached which is at $\vec{p}=(1.15,1.15)$. This behavior is similiar to ARPES measurements of the pseudogap seen in the cuprates \cite{1}. However, the pseudogap
generated by antiferromagnetic spin fluctuations disappears faster than for superconducting fluctuations. This is  due to the $\vec{q}=(\pi,\pi)$ dependence
in the the fluctuation propagator (equation (8)) which yields a 
self energy $\Sigma_{AF}( \vec{p}, \omega)$ whose magnitude is sensitive to the lack of nesting in the Fermi surface when $t^{'}$ is non-zero. 

The pseudogap region in
$A( \vec{p}, \omega )$ fills in, with the two spectral
weight peaks remaining more or less fixed in frequency $\omega$ as $\vec{p}$ is moved away from the 
{\em hot spot}. Symmetrized experimentally measured ARPES Energy Distribution Curves (EDC's) \cite{1}, similar to these model calculations, could be interpeted assuming an underlying pseudogap order parameter $\Delta_{PG}$ whose magnitude decreases to zero on an arc along the Fermi surface, along with a large
damping parameter $\Gamma$, but the approach in this work does not contain an underlying $\Delta_{PG}$, 
of course.\\
\\
In Figure 3,  $A( \vec{p}, \omega )$ for $t^{'}=0$ and $\mu=0$, the nested Fermi surface, are shown. As in
Figure 2, the left side column of panels is the superconducting fluctuation case, and the right side side is
the antiferromagnetic fluctuation case. The Fermi surface momenta $\vec{p}$ are labelled on the left column panels.
The effect of nesting is clear on the antiferromagnetic fluctuations (right side column of panels in Figure 3). A small pseudogap survives for the nodal  $\vec{p}=(1.57,1.57)$ far from the {\em hot spot} 
($(\pi,0)$), where the pseudogap effect is strongest.\\
\\
$A( \vec{p}, \omega )$ for the case of a finite frequency mode (eqns (7) and (9)),with wavevector $\vec{q}=(\pi,\pi)$ and frequency $\Omega_{0} =0.35t$ are shown in Figure 4.
In the two panels on the left of Figure 4, $t^{'}=-0.35t, \mu=-1.4t$, and
$t^{'}=0, \mu=0$ for the two right side panels.  
The role of
the {\em hot spot} in producing a peak-dip-hump structure in $A( \vec{p}, \omega )$ is clear: for the nodal point, which is $ \vec{p}=(1.15,1.15)$ (lower left panel), and $ \vec{p}=(1.57,1.57)$ (lower right panel),
this structure disappears similiar to experimental observation in some cuprates \cite{1a}.\\ 
\\
Figure 5  depicts the densities of states $N(\omega)$ (eqn. (2)) in the presence of
superconducting fluctuations (the two panels on the left), and antiferromagnetic fluctuations (the two panels on the
right). The upper panels are for $t^{'}=-0.35t, \mu=-1.4t$,and the lower panels are for $t^{'}=0, \mu=0$.
The densities of states curves in Figure 5 correspond to the same parameter values used in Figures 2 and 3. 
In the upper right panel (the antiferromagnetic fluctuation case), it is interesting to note that a pseudogap does not appear at the
Fermi energy in the density of states, even though there is one over a limited range of momenta in the spectral weight (right side column of Figure 2). Instead, a remnant of the Van Hove singularity survives.\\
\\
\\
{\bf Conclusion}\\
\\
\\
The results of this study show that, when the Fermi
surface is close to the saddle point region of the tight binding band, a
strong enhancement of the fluctuation self energy occurs for electron momenta at those regions of the Brillouin Zone. This leads to Fermi surface anisotropy
in features such as the pseudogap.

The anisotropy is similar to what is observed in high temperature
cuprates \cite{1} which display a normal state pseudogap. 

The same effect causes the Fermi surface anisotropy of the peak-dip-hump in the spectral weight
feature due to a mode. This may be the origin of the anisotropy of the peak-dip-hump in the high temperature cuprates, although this is measured
in the superconducting state in high temperature cuprates, whereas this study
is of the normal state.

\begin{figure}[h]
\begin{center}
\vspace{0.5in}
\includegraphics[scale=1.5]{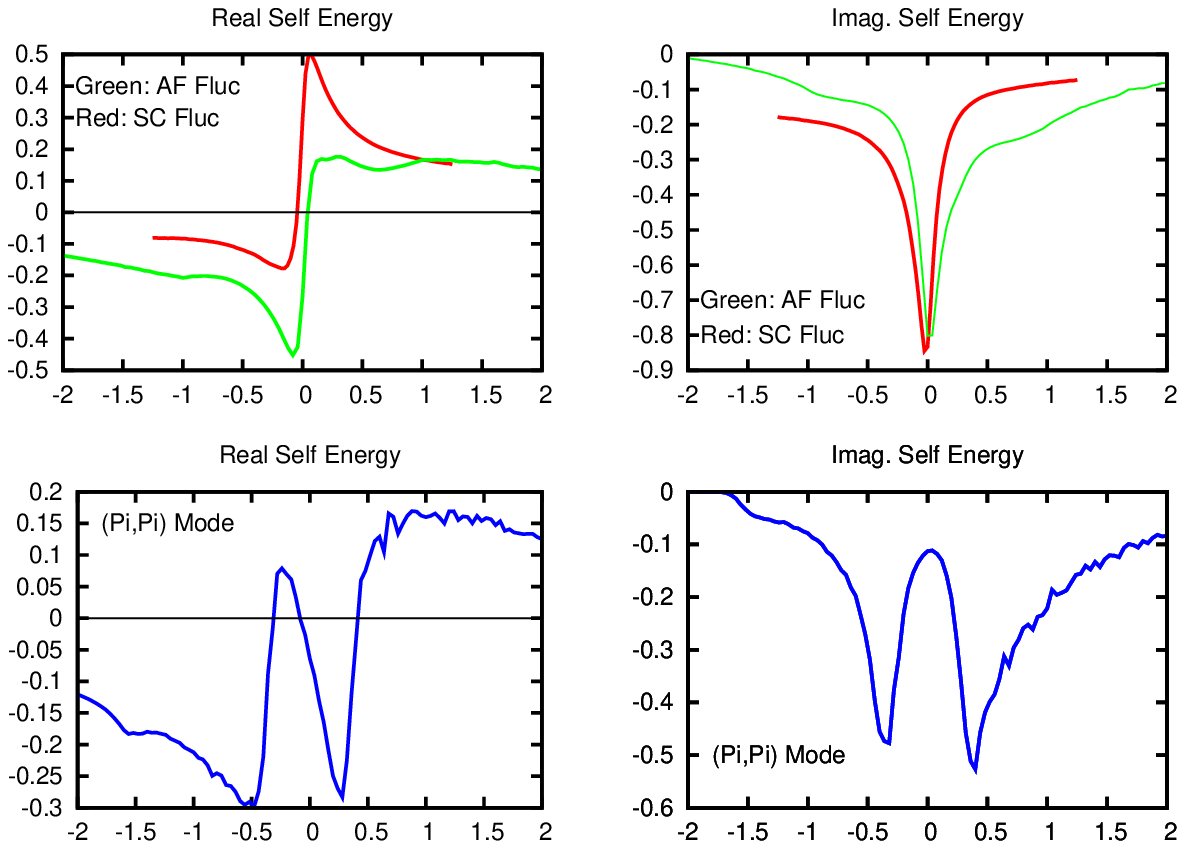}
\caption{Self Energies for superconducting fluctuations (Red),
antiferromagnetic fluctuations (Green), and a finite frequency mode (Blue) at the {\em hot spots}.
$t^{'}=-0.35t$ and $\mu=-1.4t$ in $\varepsilon_{p}$. The temperature T=0.17t. Other parameter values are listed on page 5 in the text. The horizontal and vertical axes values are $\omega/t$
and $\Sigma/t$ respectively.}
\end{center}
\end{figure}

\newpage

\begin{figure}[h]
\begin{center}
\vspace{3.5in}
\includegraphics[scale=0.85]{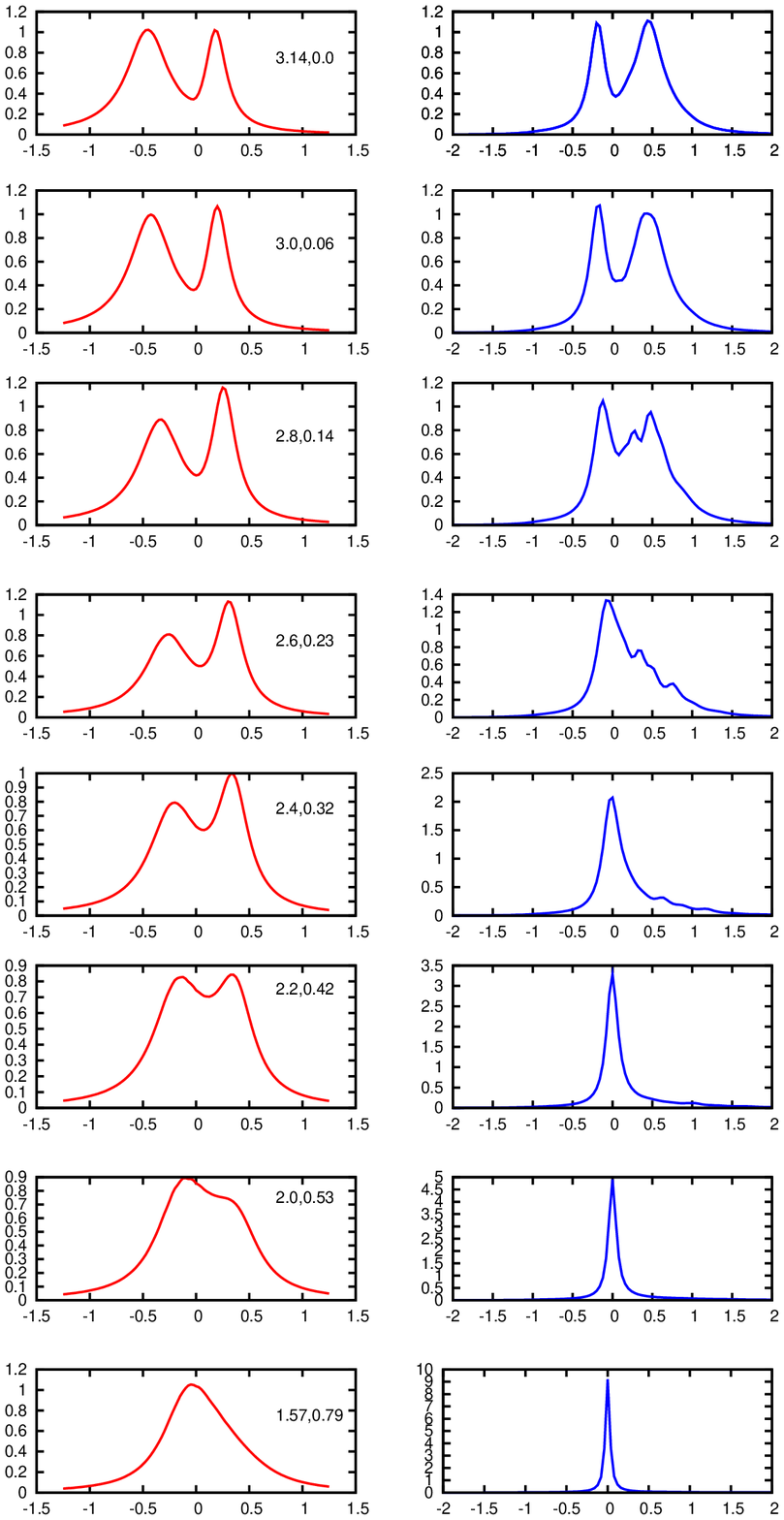}
\caption{Fermi surface spectral weight with superconducting fluctuations (left side panels), and 
antiferromagnetic fluctuations (right side panels). $t^{'}=-0.35t$ $\mu=-1.4t$ in $\varepsilon_{p}$. 
The curves start with $\vec{p}$ at the {\em hot spot} at the top, with $\vec{p}$ moving away 
from the {\em hot spot} going down the figure panels.  Horizontal axis values are $\omega/t$.}
\end{center}
\end{figure}

\newpage

\begin{figure}[h]
\begin{center}
\vspace{3.5in}
\includegraphics[scale=0.85]{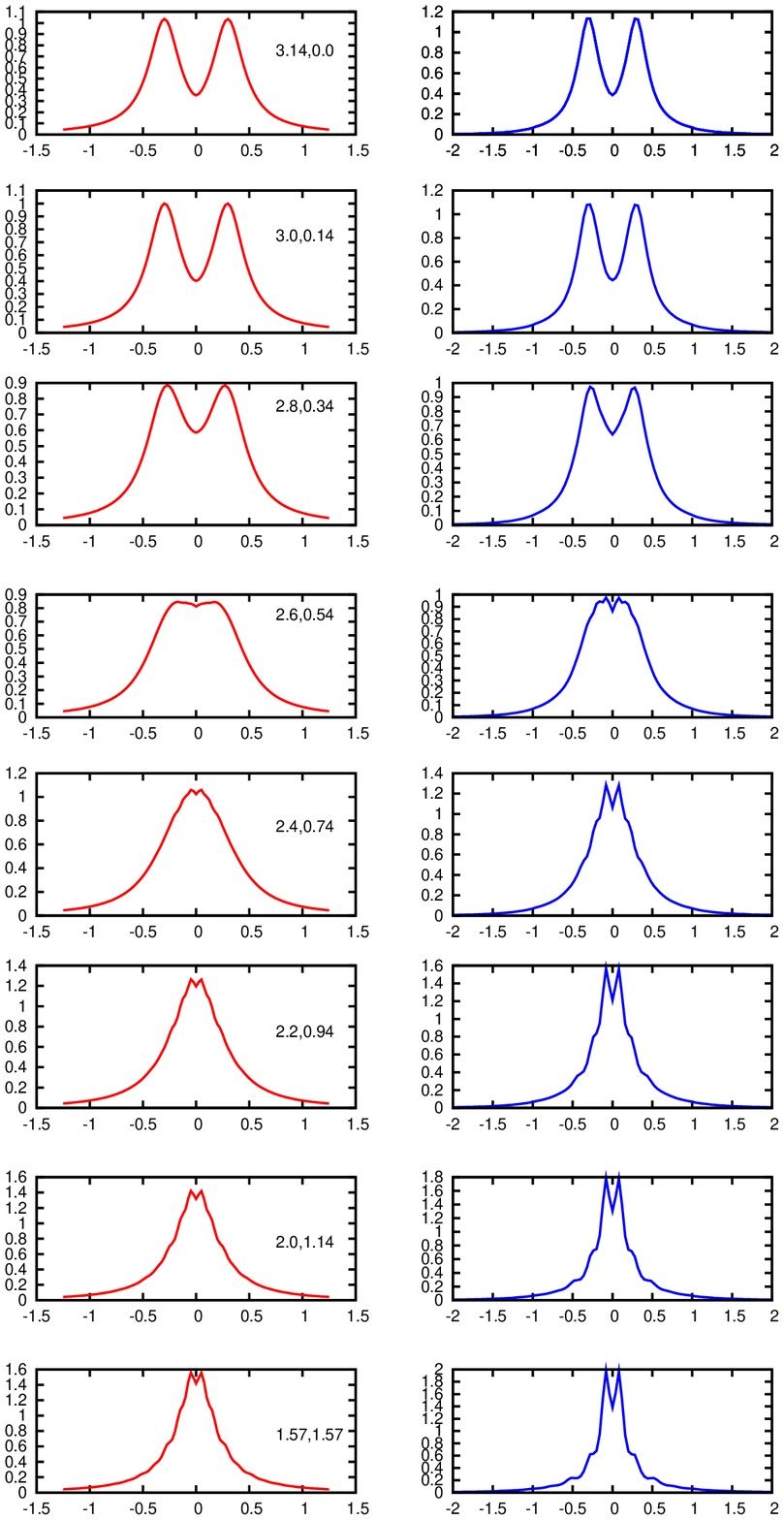}
\caption{ Fermi surface spectral weight with superconducting fluctuations (left side panels), and 
antiferromagnetic fluctuations (right side panels). $t^{'}=0$ $\mu=0$ in $\varepsilon_{p}$. 
The curves start with $\vec{p}$ at the {\em hot spot} at the top, with $\vec{p}$ moving away from the {\em hot spot} going down the figure panels.  Horizontal axis values are $\omega/t$. }
\end{center}
\end{figure}

\newpage

\begin{figure}[h]
\begin{center}
\vspace{0.5in}
\includegraphics[scale=1.5]{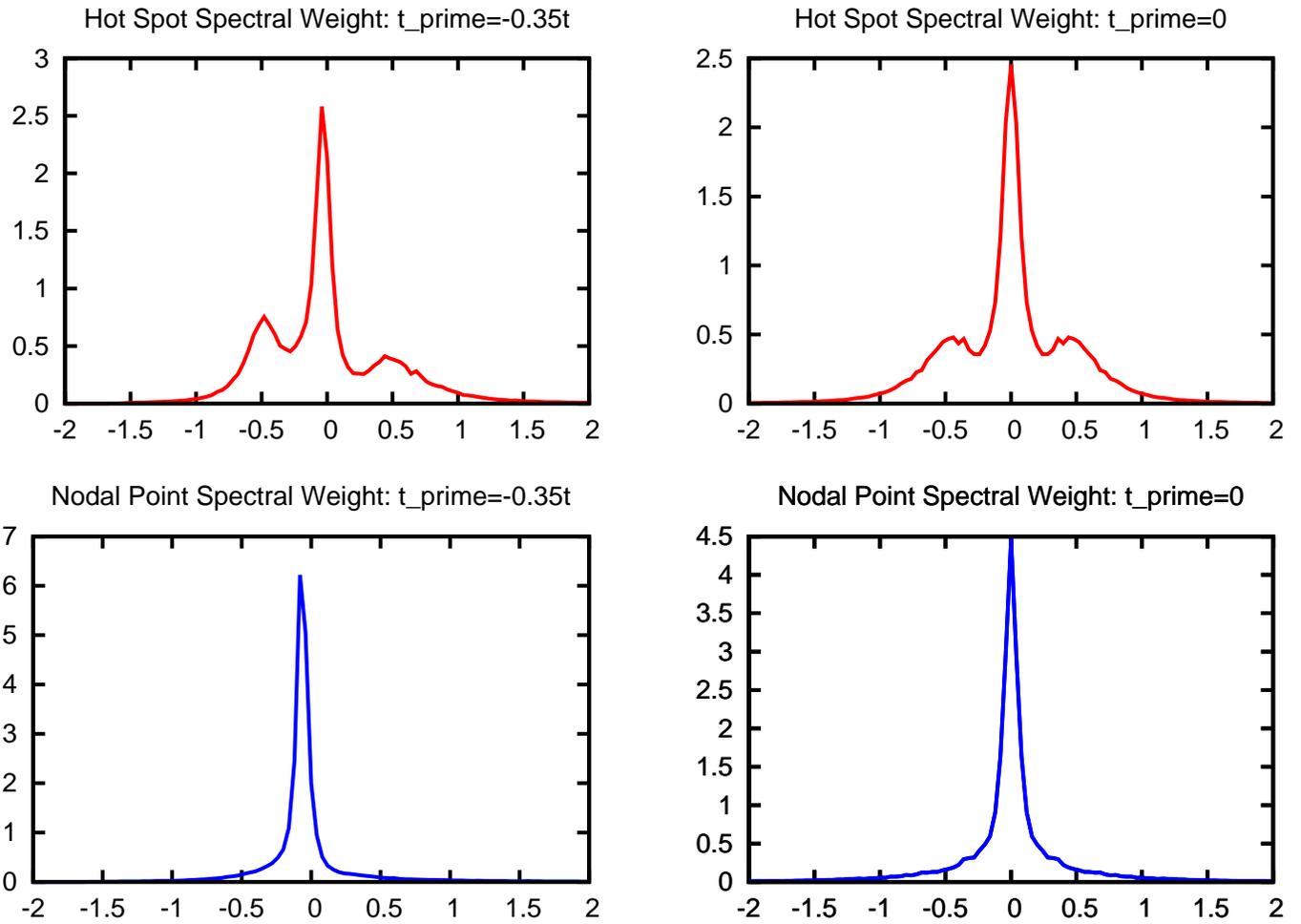}
\caption{Spectral Weights with a finite frequency mode. Left side panels for $t^{'}=-0.35t$. Right side panels
for $t^{'}=0$. Horizontal axis values are $\omega/t$.}
\end{center}
\end{figure}

\newpage

\begin{figure}[h]
\begin{center}
\vspace{0.5in}
\includegraphics[scale=1.5]{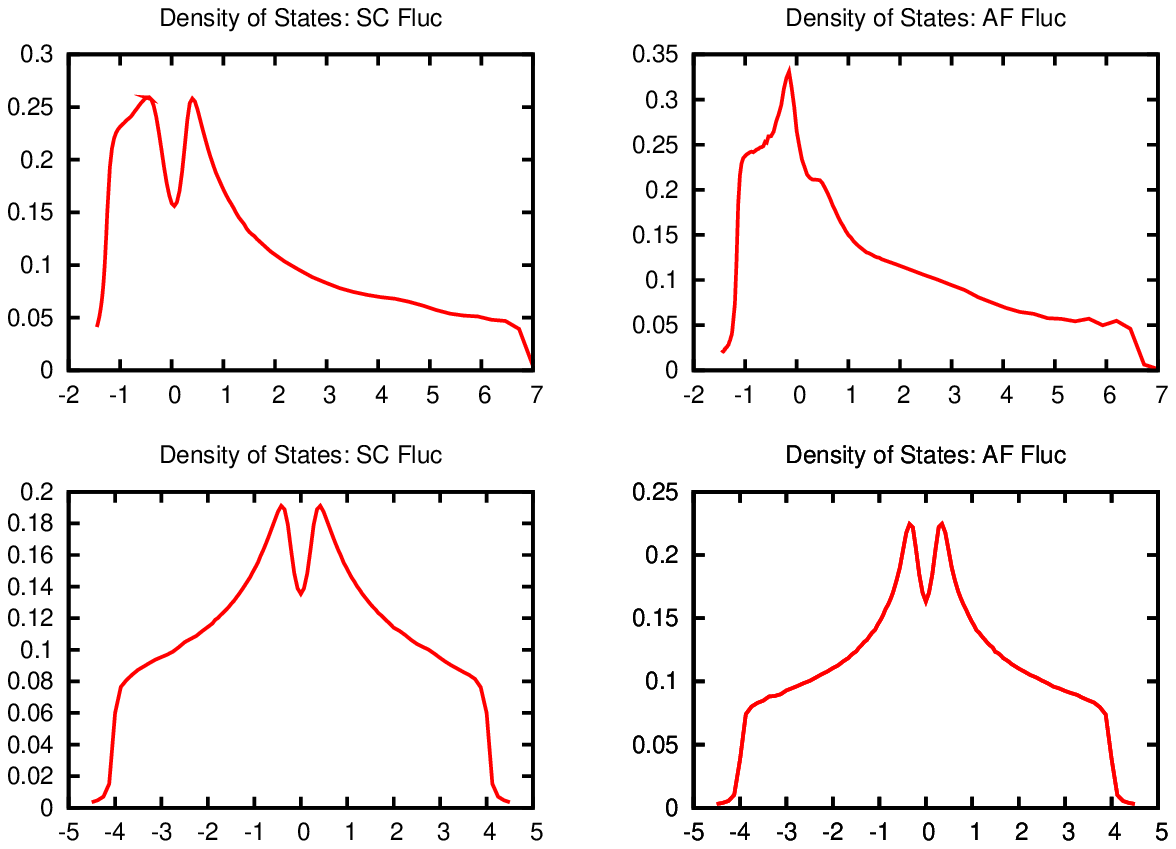}
\caption{Density of States. Upper panels 
$t^{'}=-0.35t$ $\mu=-1.4t$. Lower panels $t^{'}=0$ $\mu=0$.
Horizontal axis values are $\omega/t$.}
\end{center}
\end{figure}

\end{document}